\documentclass[preprint]{elsarticle}
\usepackage{amssymb}
\usepackage{amsmath}
\usepackage{graphicx}
\usepackage{bm}

\begin{document}

\begin{frontmatter}
\title{Two-parameter estimation with single squeezed-light interferometer via double homodyne detection}
\author[1]{Li-Li Hou}
\author[1]{Jian-Dong Zhang}
\author[2]{Kai-Min,Zheng}
\author[1]{Shuai Wang\corref{cor1}}

\cortext[cor1]{wshslxy@jsut.edu.cn}

\address[1]{School of Mathematics and Physics, Jiangsu University of Technology, Changzhou 213001, P.R. China}
\address[2]{College of Engineering and Applied Sciences, Nanjing University, Nanjing
210093, P.R. China}

\begin{abstract}
The simultaneous two-parameter estimation problem in single squeezed-light Mach-Zehnder
interferometer with double-port homodyne detection is investigated in this work.
The analytical form of the two-parameter quantum Cram\'{e}r-Bao bound defined by
the quantum Fisher information matrix is presented, which shows the ultimate limit of
the phase sensitivity will be further approved by the squeezed vacuum state.
It can not only surpass the shot-noise limit, but also can even surpass
the Heisenberg limit when half of the input intensity of the interferometer
 is provided by the coherent state and half by the squeezed light.
 For the double-port homodyne detection, the classical Fisher information matrix is
also obtained. Our results show that although the classical Cram\'{e}-Rao bound does
not saturate the quantum one, it can still asymptotically approach the
quantum Cram\'{e}r -Bao bound when the intensity of the coherent state is large enough.
Our results also indicate that the squeezed vacuum state indeed can further improve
the phase sensitivity. In addition, when half of the input intensity of the
interferometer is provided by the coherent state and half by the squeezed light,
the phase sensitivity obtained by the double-port homodyne detection can surpass
the Heisenberg limit for a small range of the estimated phase.
\end{abstract}


\end{frontmatter}

\section{Introduction}

In quantum optical metrology, it aims to reach the ultimate sensitivity
limit of the phase estimation imposed by quantum theory \cite{1,2}. Both
Mach-Zehnder interferometer (MZI) and the SU(1,1) interferometer considered
as the two-mode (two-path) interferometer have been used as a conventional
device to realize precise measurement \cite{3,4}. In general, the phase
sensitivity mainly depends on the probing quantum state of light. The
ultimate precision bound of the phase sensitivity is given by the quantum
Cram\'{e}r-Rao bound (QCRB) for any possible measurement strategy \cite{5}.
For an MZI with classical probing states, the phase sensitivity of the
estimated phase shift is bounded by the shot-noise limit (SNL) \cite{3}, $%
\Delta \phi =1/\sqrt{\bar{n}}$, where $\bar{n}$ is the mean photon number
inside the interferometer. With the help of the nonclassical quantum states,
such as squeezed states, entanglement states, it can beat the SNL, and even
approach the Heisenberg limit (HL) \cite{6,7}, $\Delta \phi =1/\bar{n}$ .\
In this case, the goal of the phase estimation is to reduce the QCRB until
it surpasses the SNL, even approaches the HL. Up to the present, extensive
theoretical and experimental research studies have been done for the
single-parameter estimation via both optical interferometers with quantum
resources \cite{8,9,10,11,12,13,14,15,16,17,18,19,20,21,22,23,24}.
Particularly, the squeezing resources is always a central strategy to
improve the phase sensitivity and has thus become a crucial concept in
quantum metrology. Recently, the quantum metrology advantage of the
squeezing resources has been also demonstrated in experiments \cite{23,24}.

In recent years, the multi-parameter estimation, i.e., the simultaneous
estimation of several parameters, has received a lot of increasing interest
\cite{25,26,27,28,29,30,31,32,33,34,35,36,37,38,39,40,r1}. It is often
necessary to estimate multiple parameters simultaneously, such as biological
system measurement \cite{41,42,43} and quantum imaging \cite{44,45,46}.
However, the current frame work of multi-parameter estimation mainly has
developed based on the quantum Fisher information matrix. As a consequence,
the phase sensitivity is\ limited by the multi-parameter QCRB which is
defined by the inverse of the quantum Fisher information matrix (QFIM). Few
measurement schemes are proposed for studying the phase sensitivity of the
multiparameter estimation. In 2014, the double-port homodyne detection has
been shown as an optimal detection for the phase and the phase diffusion
measurement for specific probe states \cite{47}. Recently, two-parameter
estimation with three-mode NOON state is investigated by the particle number
measurement \cite{48}. In this work, we will reexamine the original work by
Yurke et al \cite{4}, and show that the signal of the homodyne detection in
the MZI is not only depend on the phase difference $\phi _{d}=\phi _{1}-\phi
_{2}$, but also on the phase sum $\phi _{s}=\phi _{1}+\phi _{2}$ for some
input states. As a consequence, we propose a double-port homodyne detection
to estimate simultaneously both $\phi _{1}$ and $\phi _{2}$ in a single
squeezing-light MZI, and study its precision via the classical Fisher
information matrix (CFIM).

It is known that the single-port homodyne detection may lose almost part of
the phase information. Therefore, for single-parameter estimation, the
double-port homodyne detection has been used to improve the phase
sensitivity of the phase shift accumulated in the MZI \cite{50,51,52,53}.
The results indicate that the better phase sensitivity can be obtained by
the double-port homodyne detection \cite{53}. In addition, for quantum
illumination, a quantum receiver using the double-port homodyne detection
shows better signal-to-noise ratio and is more robust against noise \cite{54}%
. As mentioned above, it is well understood that the advantage of the
squeezed resource in the single-parameter estimation. Here, we mainly
investigate the quantum metrology advantage of the two-parameter estimation
in a squeezing-light MZI with the $\phi _{1}-\phi _{2}$ model by the
double-port homodyne detection.

The paper is organized as follows. In Sec. II, we reexamine the model and
basic theory of the MZI with two unknown phase shifts in both arms, and show
that the signal of the homodyne measurement is depend on both phases $\phi
_{d}$ and $\phi _{s}$. In Sec. III, we consider a coherent state and a
squeezed vacuum state (SVS) as the probing quantum state and investigate the
QCRB of the each phase shift $\phi _{i}$ via the QFIM. Then, we investigate
the performance of the double homodyne detection via the CFIM in Sec. IV.
Our results show that the phase sensitivity of the estimated phase can
asymptotically approach the QCRB. Finally, we summarize our results in Sec.
V.

\section{Model and basic theory}

Let us consider a balanced MZI with the $\phi _{1}-\phi _{2}$ model which
contains two 50:50 beam splitters and two phase shifts as illustrated in
Fig. 1. The first beam splitter of this device is described by the unitary
operator $\exp \left[ -i\pi J_{1}/2\right] $, and the second beam splitter
is represented by $\exp \left[ i\pi J_{1}/2\right] $. The two estimated
phase shifts $\phi _{1}$ and $\phi _{2}$ in the two arms of the MZI are
described by a phase-shifting operator, $\hat{U}\left( \phi \right) =\exp %
\left[ i\phi _{1}\hat{a}^{\dagger }\hat{a}+i\phi _{2}\hat{b}^{\dagger }\hat{b%
}\right] $. According to the SU(2) algebra of the angular momentum
operators, the overall MZI transformation reads \cite{4}
\begin{equation}
\hat{U}_{\text{MZI}}=e^{i\pi \hat{J}_{1}/2}e^{i\phi _{1}\hat{a}^{\dagger }%
\hat{a}+i\phi _{2}\hat{b}^{\dag }\hat{b}}e^{-i\pi \hat{J}_{1}/2},  \label{1}
\end{equation}%
where these angular momentum operators can be written as%
\begin{eqnarray}
\hat{J}_{1} &=&\frac{1}{2}\left( \hat{a}^{\dagger }\hat{b}+\hat{a}\hat{b}%
^{\dagger }\right) ,J_{2}=\frac{1}{2i}\left( \hat{a}^{\dagger }\hat{b}-\hat{a%
}\hat{b}^{\dagger }\right) ,  \notag \\
\text{\ }\hat{J}_{3} &=&\frac{1}{2}\left( \hat{a}^{\dagger }\hat{a}-\hat{b}%
^{\dagger }\hat{b}\right) ,\hat{J}_{0}=\frac{1}{2}\left( \hat{a}^{\dagger }%
\hat{a}+\hat{b}^{\dagger }\hat{b}\right) ,  \label{2}
\end{eqnarray}%
with the commutation relations $\left[ \hat{J}_{i},\hat{J}_{j}\right]
=i\epsilon _{ijk}\hat{J}_{k}$ ($i,j,k=1,2,3$). Here, $\hat{J}_{0}$ is a
Casimir operator that commutes with all others, i.e., $\left[ \hat{J}_{0},%
\hat{J}_{i}\right] =0$. And the operators $\hat{a}$ ($\hat{a}^{\dagger }$)
and $\hat{b}$ ($\hat{b}^{\dagger }$) are the bosonic annihilation (creation)
operators corresponding to two modes $a$ and $b$ of the interferometer,
respectively. For our purpose, according to SU(2) algebra of the angular
momentum operators and the Baker-Hausdorff lemma, Eq. (\ref{1}) can be
further recast as
\begin{equation}
\hat{U}_{\text{MZI}}=\exp \left( i\phi _{s}\hat{J}_{0}\right) \exp \left(
i\phi _{d}\hat{J}_{2}\right) ,  \label{3}
\end{equation}%
where $\phi _{s}=\phi _{1}+\phi _{2}$ represents the phase sum and $\phi
_{d}=\phi _{1}-\phi _{2\text{ }}$indicates the phase difference. Obviously,
the MZI in its full generality is a two-parameter estimation problem since
there are two unknown parameters, $\phi _{1}$ and $\phi _{2}$ (or $\phi _{d}$
and $\phi _{s}$), in the system.

\begin{figure}[tbp]
\centering\includegraphics[width=8cm]{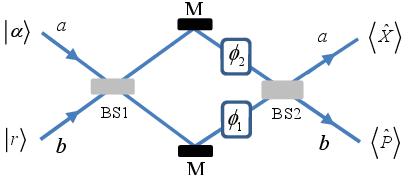}
\caption{(Color online) Scheme of the double-output homodyne detection in a
squeezed-state MZI with two-phase model.}
\end{figure}

As Yurke et al pointed \cite{4}, the operator $\exp \left[ i\phi _{s}\hat{J}%
_{0}\right] $ contained the phase sum $\phi _{s}$ commutes with particle
number operator. Therefore, in the view of the measurement, it does not
contribute to the expectation values of the number-conserving operators. For
example, for a photon-number-resolving detection, it is described by the
projection operator $\left\vert m,n\right\rangle \left\langle m,n\right\vert
$ where $\left\vert m,n\right\rangle =\hat{a}^{\dagger m}\hat{b}^{\dagger
}\left\vert 0,0\right\rangle /\sqrt{m!n!}$ is the two-mode Fock state. When
an arbitrary input $\hat{\rho}_{\text{in}}$ propagates through the MZI, the
resulted output state $\hat{\rho}_{\text{out}}$ evolves to
\begin{equation}
\hat{\rho}_{\text{out}}=\hat{U}_{\text{MZI}}\hat{\rho}_{\text{in}}\hat{U}_{%
\text{MZI}}^{\dag }.  \label{4}
\end{equation}%
Then, according to Eq. (\ref{4}), in the Heisenberg picture, the
corresponding signal Tr$\left( \hat{\rho}_{\text{out}}\left\vert
m,n\right\rangle \left\langle m,n\right\vert \right) $ of the
photon-number-resolving detection reads
\begin{equation}
\text{Tr}\left\{ \hat{\rho}_{\text{in}}\exp \left( -i\phi _{d}\hat{J}%
_{y}\right) \left\vert m,n\right\rangle \left\langle m,n\right\vert \exp
\left( i\phi _{d}\hat{J}_{y}\right) \right\} .  \label{e1}
\end{equation}%
Obviously, the operator $\exp \left( i\phi _{s}\hat{J}_{0}\right) $
contained the phase sum factor does not contribute to the expectation values
of the photon-number-resolving detection for an arbitrary input state. In
deriving Eq. (\ref{e1}), the following the transformations $\exp \left(
-i\phi _{s}\hat{J}_{0}\right) \hat{a}^{\dagger }\exp \left( i\phi _{s}\hat{J}%
_{0}\right) =\hat{a}^{\dagger }\exp \left( -i\phi _{s}/2\right) $ and $\exp
\left( -i\phi _{s}\hat{J}_{0}\right) \hat{b}^{\dag }\exp \left( i\phi _{s}%
\hat{J}_{0}\right) =\hat{b}^{\dag }\exp \left( -i\phi _{s}/2\right) $ have
been used. It can be seen that, for intensity detection or parity detection
based on the photon-number-resolving detector, including the
photon-number-resolving detection, the signals of these measurements are
only depended on the phase difference $\phi _{d}$. As a consequence, in the
past decades, one is usually interested in the phase difference contained in
the MZI. Especially, one main consider two phase-shift configurations: one
is the phase anti-symmetrically distribution, i.e., $\phi _{1}=-\phi /2$ and
$\phi _{2}=\phi /2$, another is the phase shift only occurs in one mode of
the interferometer (for example $\phi _{1}=\phi $ and $\phi _{2}=0$).

However, according to the characteristics of the structure of the MZI
expressed by Eq. (\ref{3}), the signal of the homodyne detection is usually
sensitive to both $\phi _{d}$ and $\phi _{s}$ for some input states.
Therefore, we need consider both QFIM and CFIM to investigate the phase
sensitivity. In the following, we will investigate the two-parameter
estimation in a single MZI with the $\phi _{1}-\phi _{2}$ model via the
double-port homodyne detection.

\section{The QCRB of two-parameter estimation}

In the multi-parameter estimation, we need introduce the multi-parameter
QCRB, which is inversely proportional to the quantum Fisher information,
i.e., the bound is given by

\begin{equation}
\sum_{i}\left( \Delta ^{2}\phi _{i}\right) _{\text{QCRB}}=\text{Tr}\left[
\left( F^{q}\right) ^{-1}\right] ,  \label{6}
\end{equation}%
where $\sum_{i}\left( \Delta ^{2}\phi _{i}\right) _{\text{QCRB}}$ is the
total variance for multi-parameter estimation to quantity the measurement
precision, and $\left( F^{q}\right) ^{-1}$ is the inverse of the QFIM. In
the MZI with the $\phi _{1}-\phi _{2}$ model, the matrix elements $%
F_{ij}^{q} $ of the two-by-two QFIM is defined as
\begin{equation}
F_{ij}^{q}=4\mathrm{Re}\left[ \left\langle \partial _{i}\psi \left( \phi
\right) \right\vert \left\vert \partial _{j}\psi \left( \phi \right)
\right\rangle -\left\langle \partial _{i}\psi \left( \phi \right)
\right\vert \left\vert \psi \left( \phi \right) \right\rangle \left\langle
\psi \left( \phi \right) \right\vert \left\vert \partial _{j}\psi \left(
\phi \right) \right\rangle \right] ,  \label{7}
\end{equation}%
with $\left\vert \psi \left( \phi \right) \right\rangle =\hat{U}\left( \phi
\right) \exp \left[ -i\pi J_{x}/2\right] \left\vert \text{in}\right\rangle $
is the probing state after sensing both phase shifts and $\left\vert
\partial _{i}\psi \left( \phi \right) \right\rangle =\partial \left\vert
\psi \left( \phi \right) \right\rangle /\partial \phi _{i}$. The subscript "$%
i$" and "$j$" denote the $1$ and $2$, respectively. It can be seen that the
QFIM is solely determined by the parameterized output state and its
dependence on the parameters. In addition, it is well known that the QCRB
defined by the QFIM gives an ultimate bound of the phase sensitivity over
all possible measurements.

Here, we consider a coherent state and an SVS, $\left\vert \text{in}%
\right\rangle =\left\vert z\right\rangle _{a}\left\vert r\right\rangle _{b}$%
, as the probing state of the MZI. The SVS is defined by $\left\vert
r\right\rangle =\mathrm{sech}^{1/2}r\exp \left[ -\hat{b}^{\dagger 2}\tanh
re^{i\theta _{s}}/2\right] \left\vert 0\right\rangle $. The average total
photon number of the input state is given by $\bar{n}_{\text{in}}=\left\vert
z\right\vert ^{2}+\sinh ^{2}r$. For the convenience, we rewrite the SVS in
the basis of the coherent state
\begin{equation}
\left\vert r\right\rangle =\mathrm{sech}^{1/2}r\int \frac{d^{2}\beta }{\pi }%
e^{-\frac{1}{2}\left\vert \beta \right\vert ^{2}-\frac{\tanh r}{2}e^{i\theta
_{s}}\beta ^{\ast 2}}\left\vert \beta \right\rangle ,  \label{a6}
\end{equation}%
where $\left\vert \beta \right\rangle $ represents a coherent state. Then,
we can further obtain the expectation of the operator $\hat{b}^{\dag n}\hat{b%
}^{m}$ in the SVS
\begin{equation}
\left\langle r\right\vert \hat{b}^{\dag n}\hat{b}^{m}\left\vert
r\right\rangle =\frac{\partial ^{m+n}}{\partial t^{m}\partial \tau ^{n}}\exp %
\left[ -\left( t^{2}e^{-i\theta _{s}}+\tau ^{2}e^{i\theta _{s}}\right) \frac{%
\sinh 2r}{4}+\tau t\sinh ^{2}r\right] _{t,\tau =0}.  \label{8}
\end{equation}%
Given the probe state, we can now calculate the matrix elements of the QFIM.
According to Eqs. (\ref{7}) and \ref{8}, we obtain the corresponding QFIM,
i.e.,%
\begin{equation}
F^{q}=\left[
\begin{array}{cc}
F_{11}^{q} & F_{12}^{q} \\
F_{21}^{q} & F_{22}^{q}%
\end{array}%
\right] ,  \label{9}
\end{equation}%
with the matrix elements
\begin{equation}
F_{11}^{q}=F_{22}^{q}=\left( 2\sinh ^{2}r+3\right) \sinh ^{2}r+\left\vert
z\right\vert ^{2}\left( 1+e^{2r}\right) ,  \label{10}
\end{equation}%
and%
\begin{equation}
F_{12}^{q}=F_{21}^{q}=\left( 2\sinh ^{2}r+1\right) \sinh ^{2}r+\left\vert
z\right\vert ^{2}\left( 1-e^{2r}\right) .  \label{11}
\end{equation}%
In order to maximize the QCRB, we have adopted the combined phase $2\psi
+\theta _{s}=2k\pi $ ($k=0,1,2\cdots $). The factor $\psi $ is the phase of
the complex amplitude $z$ of the coherent state $\left\vert z\right\rangle $%
. In the MZI with the $\phi _{1}-\phi _{2}$ model, we mainly focus on the
estimation of each phase shift $\phi _{i}$. For $\phi _{1}$ or $\phi _{2}$,
it can be seen from Eqs. (\ref{9}-\ref{11}) that the QCRB is the same, i.e.,
the ultimate bound of the phase sensitivity of the phase shift $\phi _{1}$
or $\phi _{2}$ reads

\begin{eqnarray}
\left( \Delta ^{2}\phi _{i}\right) _{\text{QCRB}} &=&\frac{F_{ii}^{q}}{%
\left( F_{ii}^{q}\right) ^{2}-\left( F_{i,j}^{q}\right) ^{2}}  \notag \\
&=&\frac{\left( 2\sinh ^{2}r+3\right) \sinh ^{2}r+\left\vert z\right\vert
^{2}\left( 1+e^{2r}\right) }{2\left( \sinh ^{2}r+\left\vert z\right\vert
^{2}e^{2r}\right) \left( \sinh ^{2}2r+2\left\vert z\right\vert ^{2}\right) }.
\label{12}
\end{eqnarray}%
These results expressed by Eqs. (\ref{9}-\ref{11}) are somewhat different
from that results expressed by Eq. (10) in Ref. \cite{55}. This is because
that here we calculate the QFIM in basis $\phi _{1}$ and $\phi _{2}$, i.e.,
the MZI with the $\phi _{1}-\phi _{2}$ model. While, QFIM is obtained in
basis $\phi _{d}$ and $\phi _{s}$ in Ref. \cite{55}. Particularly, in the
case of $r=0$, Eq. (\ref{12}) reduces to $\left( \Delta ^{2}\phi _{i}\right)
_{\text{QCRB}}|_{r=0}=1/\left( 2\left\vert z\right\vert ^{2}\right) $.
Actually, for the MZI with the $\phi _{1}-\phi _{2}$ model, the QCRB of each
phase shift $\phi _{i}$ can still surpass the SNL, even one mode of the
interferometer is a vacuum state as shown in Eq. (\ref{12}). It is
interesting that, in the case of the $\left\vert z\right\vert ^{2}=\sinh
^{2}r=\bar{n}_{\text{in}}/2$, Eq. (\ref{12}) reduces to

\begin{equation}
\left( \Delta ^{2}\phi _{i}\right) _{\text{QCRB}}=\frac{\left( 3+\bar{n}_{%
\text{in}}\right) +\left( 2+\bar{n}_{\text{in}}+\sqrt{\bar{n}_{\text{in}%
}\left( 2+\bar{n}_{\text{in}}\right) }\right) }{2\bar{n}_{\text{in}}\left( 2+%
\bar{n}_{\text{in}}+\sqrt{\bar{n}_{\text{in}}\left( 2+\bar{n}_{\text{in}%
}\right) }\right) \left( 3+\bar{n}_{\text{in}}\right) }.  \label{a0}
\end{equation}%
which can surpass the HL. At the same time, the QCRB reaches its minimum in
the case of $\left\vert z\right\vert ^{2}=\sinh ^{2}r$ for a give mean
photon number of the input state. On the other hand, according to Eq. (\ref%
{10}) when the phase shift is generated only in one mode of the MZI, the
QCRB can described by $1/F_{ii}^{q}$, i.e.,%
\begin{equation}
\left( \Delta ^{2}\phi \right) _{\text{QCRB}}=\frac{1}{\left( 2\sinh
^{2}r+3\right) \sinh ^{2}r+\left\vert z\right\vert ^{2}\left(
1+e^{2r}\right) },  \label{w1}
\end{equation}%
which is consistent with the Eq. (4) in Ref.\cite{55}.

\begin{figure}[tbp]
\centering\includegraphics[width=8cm]{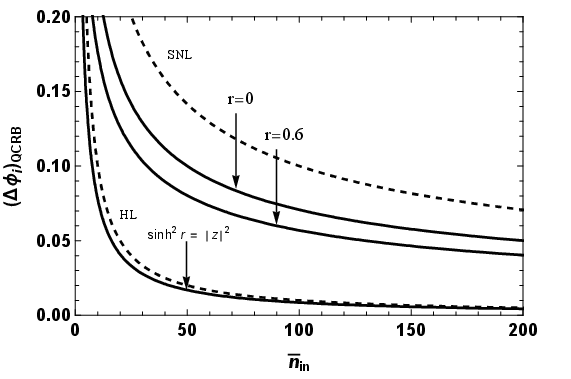}
\caption{(Color online) The QCRB of the phase $\protect\phi _{i}$ as a
function of the total mean photon number inside the interferometer for given
$r=0.0$ and $r=0.6$. The upper black dashed line denotes the SNL, while the
below black dashed line represents the HL.}
\end{figure}

In order to illustrate the Eq. (\ref{12}) explicitly and more exactly, we
plot the QCRB with respect to the average total photon number $\bar{n}_{%
\text{in}}$ in Fig. 2. It can be seen that the SVS can further improve the
QCRB. In addition, in the case of the $\left\vert z\right\vert ^{2}=\sinh
^{2}r=\bar{n}_{\text{in}}/2$, the QCRB can even surpass the HL. It may be an
interesting result. It should be noted that, for the single-parameter
estimation, very recently it has been shown in experiment that the phase
sensitivity obtained by the double-port homodyne detection with an SVS as a
probe can surpass the HL as well as the ideal N00N state protocol \cite{23}.

\section{Two-parameter estimation with single squeezed-light MZI via
double-port homodyne detection}

As mentioned above, for the MZI with two estimated phase shifts, the signal
of the the parity detection, intensity detection and photon-number-resolving
detection is only sensitive to the phase difference. However, for the
homodyne detection, things are somewhat different. Here, we first
demonstrate analytically that the signal of the homodyne detection is
sensitive to both $\phi _{d}$ and $\phi _{s}$ for a coherent state and an
SVS considered as the probe state. Then, we consider the simultaneous
estimation of the two phase shifts $\phi _{1}$ and $\phi _{2}$ in a
squeezing-light MZI via double-port homodyne detection, and study its
precision based on the CFIM.

\subsection{The signal of the double-port homodyne detection}

The homodyne detection is modeled as projection on quadratures of the
bosonic filed described by the standard quadrature operator \cite{56}, i.e.,
\begin{equation}
\hat{X}\left( \theta \right) =\frac{\hat{a}e^{-i\theta }+\hat{a}^{\dagger
}e^{i\theta }}{\sqrt{2}},  \label{13}
\end{equation}%
where $\theta $ is a phase of the strong local oscillator. Here, we consider
the double-port homodyne detection as shown in Fig. 1. The two outports of
the interferometer are projected on the $\hat{X}_{a}=\hat{X}\left( 0\right) $
and $\hat{P}_{b}=\hat{X}\left( \pi /2\right) $ quadratures via the homodyne
measurement. Here, the subscripts "$a$" and "$b$" denote two different field
modes $a$ and $b$ of the interferometer. According to Eq. (\ref{4}), for an
arbitrary input state $\hat{\rho}_{\text{in}}$, the signal of the homodyne
detection in the Heisenberg picture reads
\begin{equation}
\left\langle \hat{X}\left( \theta \right) \right\rangle =\text{Tr}\left\{
\hat{\rho}_{\text{in}}\hat{U}_{\text{MZI}}^{\dagger }\hat{X}\left( \theta
\right) \hat{U}_{\text{MZI}}\right\} .  \label{14}
\end{equation}%
Then, noting that the following field operator transformations of the MZI%
\begin{eqnarray}
\hat{U}_{\text{MZI}}^{\dagger }\hat{a}\hat{U}_{\text{MZI}} &=&\left( \hat{a}%
\cos \frac{\phi _{d}}{2}+\hat{b}\sin \frac{\phi _{d}}{2}\right) \exp \left( i%
\frac{\phi _{s}}{2}\right) ,  \notag \\
\hat{U}_{\text{MZI}}^{\dagger }\hat{b}\hat{U}_{\text{MZI}} &=&\left( -\hat{a}%
\sin \frac{\phi _{d}}{2}+\hat{b}\cos \frac{\phi _{d}}{2}\right) \exp \left( i%
\frac{\phi _{s}}{2}\right) ,  \label{e2}
\end{eqnarray}%
one can clearly see that the signal of the homodyne detection, in general,
depends on both $\phi _{d}$ and $\phi _{s}$ for the input state $\hat{\rho}_{%
\text{in}}$. Enlighten by this result, in the following we will investigate
the two-parameter estimation with a squeezing-light MZI via the double-port
homodyne detection. When we consider a coherent state and an SVS, $\hat{\rho}%
_{\text{in}}=\left\vert z\right\rangle _{a}\left\vert r\right\rangle
_{b}\left. _{b}\left\langle r\right\vert _{a}\left\langle z\right\vert
\right. $, as the probing state of the MZI, according to Eq. (\ref{14}), we
can immediately obtain the signals of the double-port homodyne detection,
i.e.,

\begin{equation}
\left\langle \hat{X}_{a}\right\rangle =\sqrt{2}\left\vert z\right\vert \cos
\frac{\phi _{d}}{2}\cos \left( \psi +\frac{\phi _{s}}{2}\right) ,  \label{a1}
\end{equation}%
and%
\begin{equation}
\left\langle \hat{P}_{b}\right\rangle =-\sqrt{2}\left\vert z\right\vert \sin
\frac{\phi _{d}}{2}\sin \left( \psi +\frac{\phi _{s}}{2}\right) ,  \label{a2}
\end{equation}%
respectively. One can clearly see that the signal of the homodyne detection
depends on both the phase difference and phase sum for an such probe state.
As a consequence, one can simultaneously obtain the values of both phases $%
\phi _{1}$ and $\phi _{2}$ via Eqs. (\ref{a1}) and (\ref{a2}). In addition,
the different phase-configuration will lead to the different signals of the
measurement, even for single-parameter estimation. For example, if we are
only interested in the single phase difference $\phi _{d}$, we still will
obtain different values of the quantum Fisher information when we choice
different phase-configurations \cite{r2,r3}, e.g., the phase-configuration
with $\phi _{1}=-\phi /2$ and $\phi _{2}=\phi /2,$ or the case of $\phi
_{1}=\phi $ and $\phi _{2}=0$ (or $\phi _{1}=0,\phi _{2}=\phi $ ). This
result is also occurred in the phase estimation of the SU(1,1)
interferometer \cite{r4,r5}. Therefore, this two-parameter model of the MZI,
as shown in Eq. (\ref{3}), explicitly shows that one does not know the phase
$\phi _{s}$ in prior. This will affect the precision limit of the phase sum $%
\phi _{d}$ especially when $\phi _{d}$ and $\phi _{s}$ are correlated. As a
consequence, we need calculate the two-by-two CFIM with respect to the
double homodyne detection ($\hat{X}$, $\hat{P}$) to access the measurement
precision.

\subsection{Sensitivity determined by the classical Fisher information matrix%
}

\bigskip The Fisher information lies at the heart of the phase sensitivity
of the estimated phases. In quantum metrology, the CRB is determined by the
CFIM and sets the ultimate limit bound of the phase sensitivity with a
specific measurement strategy. For the MZI with the $\phi _{1}-\phi _{2}$
model, its CFIM is given by a two-by-two matrix \cite{57}%
\begin{equation}
F^{c}=\left[
\begin{array}{cc}
F_{11}^{c} & F_{12}^{c} \\
F_{21}^{c} & F_{22}^{c}%
\end{array}%
\right] ,  \label{a3}
\end{equation}%
where the subscript $1$ and $2$ correspond to $\phi _{1}$ and $\phi _{2}$.
After a sequence of detection events, one can obtain the two-parameter CRB
to quantify the phase sensitivity, i.e.,
\begin{equation}
\sum_{i}\left( \Delta ^{2}\phi _{i}\right) =\text{Tr}\left[ \left(
F^{c}\right) ^{-1}\right] ,  \label{a4}
\end{equation}%
Therefore, the CRB of the two-parameter estimation with a given measurement
is determined by the Tr$\left[ \left( F^{c}\right) ^{-1}\right] $. Because
the results ($x_{a},p_{b}$) of the measurement ($\hat{X}_{a}$, $\hat{P}_{b}$%
) are continuous variables, the elements of the CFIM should be expressed in
the integral form

\begin{equation}
F_{i,j}^{c}=\int \frac{dx_{a}dp_{b}}{p\left( x_{a},p_{b}|\phi _{1},\phi
_{2}\right) }\frac{\partial p\left( x_{a},p_{b}|\phi _{1},\phi _{2}\right) }{%
\partial \phi _{i}}\frac{\partial p\left( x_{a},p_{b}|\phi _{1},\phi
_{2}\right) }{\partial \phi _{j}}  \label{a5}
\end{equation}%
where the subscript "$i$" and "$j$" denote $1$ and $2$, and $p\left(
x_{a},p_{b}|\phi _{1},\phi _{2}\right) $ is the phase-dependent probability
distribution of the measurement results ($x_{a},p_{b}$). For the
double-homodyne detection, the conditional probability function $p\left(
x_{a},p_{b}|\phi _{1},\phi _{2}\right) $ is just the expectation value of
the projection operator $\left\vert x_{a},p_{b}\right\rangle \left\langle
x_{a},p_{b}\right\vert $ in the output state, i.e.,
\begin{equation}
p\left( x_{a},p_{b}\right) |\left( \phi _{s},\phi _{d}\right) =\text{Tr}%
\left[ \hat{U}_{\text{MZI}}\hat{\rho}_{\text{in}}\hat{U}_{\text{MZI}}^{\dag
}\left\vert x_{a},p_{b}\right\rangle \left\langle x_{a},p_{b}\right\vert %
\right] ,  \label{15}
\end{equation}%
where the coordinate and momentum eigenstate $\left\vert
x_{a},p_{b}\right\rangle $ reads \cite{58}%
\begin{eqnarray}
\left\vert x_{a},p_{b}\right\rangle  &=&\left( \frac{1}{\pi }\right)
^{1/2}\exp \left[ -\frac{x_{a}^{2}}{2}+\sqrt{2}x_{a}\hat{a}^{\dag }-\frac{%
\hat{a}^{\dag 2}}{2}\right] \left\vert 0\right\rangle _{a}  \notag \\
&&\otimes \exp \left[ -\frac{p_{b}^{2}}{2}+i\sqrt{2}p_{b}\hat{b}^{\dag }+%
\frac{\hat{b}^{\dag 2}}{2}\right] \left\vert 0\right\rangle _{b}.  \label{16}
\end{eqnarray}%
For the input state $\hat{\rho}_{\text{in}}=\left\vert z\right\rangle
_{a}\left\vert r\right\rangle _{b}\left. _{b}\left\langle r\right\vert
_{a}\left\langle z\right\vert \right. $, according to Eq. (\ref{15}), after
long but straightforward calculation, we can obtain the probability of the
outcome ($x_{a},p_{b}$)
\begin{eqnarray}
P\left( x_{a},p_{b}\right) |\left( \phi _{1},\phi _{2}\right)  &=&\frac{A_{0}%
}{\pi }\exp \left[ A_{1}x_{a}+B_{1}p_{b}\right.   \notag \\
&&\left. -A_{2}x_{a}^{2}-B_{2}p_{b}^{2}+C_{0}x_{a}p_{b}+D_{0}\right] ,
\label{17}
\end{eqnarray}%
where
\begin{eqnarray}
A_{0} &=&\left[ \cosh ^{2}r+\cos \phi _{d}\cos \left( \phi _{s}+\theta
_{s}\right) \sinh 2r+\cos ^{2}\phi _{d}\sinh ^{2}r\right] ^{-1/2},  \notag \\
C_{0} &=&-A_{0}^{2}\sin \phi _{d}\sin \left( \phi _{s}+\theta _{s}\right)
\sinh 2r,  \notag \\
D_{0} &=&-\left\vert z\right\vert ^{2}\left[ \cos \phi _{d}\cos \left( \phi
_{s}+2\psi \right) \left( 1+A_{0}^{2}\sin ^{2}\phi _{d}\sinh ^{2}r\right)
\right.   \notag \\
&&\left. +A_{0}^{2}\sin ^{2}\phi _{d}\cos \left( 2\phi _{s}+\theta
_{s}+2\psi \right) \sinh r\cosh r\right] ,  \label{a7}
\end{eqnarray}%
and
\begin{eqnarray}
A_{1} &=&2\sqrt{2}\left\vert z\right\vert \left[ \cos \frac{\phi _{d}}{2}%
\cos \left( \frac{\phi _{s}}{2}+\psi \right) \right.   \notag \\
&&\left. +\frac{1}{2}A_{0}^{2}\sin \frac{\phi _{d}}{2}\sin 2\phi _{d}\cos
\left( \frac{\phi _{s}}{2}+\psi \right) \sinh ^{2}r\right.   \notag \\
&&\left. +A_{0}^{2}\sin \frac{\phi _{d}}{2}\sin \phi _{d}\cos \left( \frac{3%
}{2}\phi _{s}+\theta _{s}+\psi \right) \sinh r\cosh r\right] ,  \notag \\
B_{1} &=&2\sqrt{2}\left\vert z\right\vert \left[ -\sin \frac{\phi _{d}}{2}%
\sin \left( \frac{\phi _{s}}{2}+\psi \right) \right.   \notag \\
&&\left. +\frac{1}{2}A_{0}^{2}\cos \frac{\phi _{d}}{2}\sin 2\phi _{d}\sin
\left( \frac{\phi _{s}}{2}+\psi \right) \sinh ^{2}r\right.   \notag \\
&&\left. +A_{0}^{2}\cos \frac{\phi _{d}}{2}\sin \phi _{d}\sin \left( \frac{3%
}{2}\phi _{s}+\theta _{s}+\psi \right) \sinh r\cosh r\right] ,  \label{a8}
\end{eqnarray}%
as well as%
\begin{eqnarray}
A_{2} &=&1+2A_{0}^{2}\sin ^{2}\frac{\phi _{d}}{2}\cos \phi _{d}\sinh ^{2}r
\notag \\
&&+2A_{0}^{2}\sin ^{2}\frac{\phi _{d}}{2}\cos \left( \phi _{s}+\theta
_{s}\right) \sinh r\cosh r,  \notag \\
B_{2} &=&1-2A_{0}^{2}\cos ^{2}\frac{\phi _{d}}{2}\cos \phi _{d}\sinh ^{2}r
\notag \\
&&-2A_{0}^{2}\cos ^{2}\frac{\phi _{d}}{2}\cos \left( \theta _{s}+\phi
_{s}\right) \sinh r\cosh r,  \label{a9}
\end{eqnarray}%
with the relation $4A_{2}B_{2}-C_{0}^{2}=4A_{0}^{2}$. Noting that the
integral formula
\begin{equation}
\int e^{-\alpha x^{2}+\beta x}dx=\sqrt{\frac{\pi }{\alpha }}e^{\frac{\beta
^{2}}{4\alpha }},\mathrm{Re}\left( \alpha \right) >0,  \label{19}
\end{equation}%
one can demonstrate that $\int dx_{a}dp_{b}P\left( x_{a},p_{b}\right)
|\left( \phi _{1},\phi _{2}\right) =1$ is satisfied. In order to obtain the
analytical expressions of the elements of the FIM, we also need obtain the
expected value of the operator $\hat{X}_{a}^{m}\hat{P}_{b}^{n}$ in the
output state. Based on Eq. (\ref{17}), we can directly derive such
expectation value as the following
\begin{eqnarray}
&&\text{Tr}\left( \hat{\rho}_{\text{out}}\hat{X}_{a}^{m}\hat{P}%
_{b}^{n}\right)   \notag \\
&=&\int x_{a}^{m}p_{b}^{n}dx_{1}dp_{2}P\left( x_{a},p_{b}\right) |\left(
\phi _{1},\phi _{2}\right)   \notag \\
&=&\frac{A_{0}\partial ^{m+n}}{\pi \partial t^{m}\partial \tau ^{n}}\int
dx_{a}dp_{b}e^{\left( A_{1}+t\right) x_{a}+\left( B_{1}+\tau \right)
p_{b}-A_{2}x_{a}^{2}-B_{2}p_{b}^{2}+Cx_{a}p_{b}+D}|_{t,\tau =0}  \label{20}
\end{eqnarray}%
Then, after integration, we can immediately obtain%
\begin{equation}
\text{Tr}\left( \hat{\rho}_{\text{out}}\hat{X}_{a}^{m}\hat{P}_{b}^{n}\right)
=\frac{\partial ^{m+n}}{\partial t^{m}\partial \tau ^{n}}e^{\frac{B_{2}}{%
4A_{0}^{2}}\left( A_{1}+\frac{B_{1}C+C\tau }{2B_{2}}+t\right) ^{2}+\frac{%
\left( \tau +B_{1}\right) ^{2}}{4B_{2}}+D}|_{t,\tau =0}  \label{21}
\end{equation}%
In the case of $\left( m=1\text{, }n=0\right) $ and $\left( m=0\text{, }%
n=1\right) $, we can re-obtain Eqs. (\ref{a1}) and (\ref{a2}) by Eq. (\ref%
{21}), respectively. Combining Eqs. (\ref{a5}) and (\ref{17}), we can
finally obtain the elements of the CFIM. As a consequence, for each
estimated phase shift $\phi _{1}$ or $\phi _{2}$, we can obtain its phase
sensitivity, i.e.,

\begin{equation}
\Delta ^{2}\phi _{i}=\frac{F_{ii}^{c}}{F_{ii}^{c}F_{jj}^{c}-\left(
F_{i,j}^{c}\right) ^{2}},  \label{22}
\end{equation}%
In general, $F_{11}^{c}\neq F_{22}^{c}$ for any values of both $\phi _{1}$
and $\phi _{2}$. The analytical expressions of the elements of the CFIM are
very tedious and lengthy. Here, we do not show them in the text. In order to
optimized the CRB, we set $\psi =0.5\pi $ and $\theta _{s}=\pi $ in the
following.

In the case of $\psi =0.5\pi $ and $\theta _{s}=\pi $, we can obtain the
concise expressions of the CFIM at $\phi _{1}=\phi _{2}=0$ as the following
\begin{equation}
F^{c}=\left\vert z\right\vert ^{2}\left[
\begin{array}{cc}
1+e^{2r} & 1-e^{2r} \\
1-e^{2r} & 1+e^{2r}%
\end{array}%
\right] .  \label{24}
\end{equation}%
It can be seen that $F_{11}^{c}=F_{22}^{c}$ in the case of $\phi _{1}=\phi
_{2}=0$. According to Eqs. (\ref{24}), for the MZI with the $\phi _{1}-\phi
_{2}$ model, it can be also seen that the phase sensitivity of both phase
shift $\phi _{1}$ or $\phi _{2}$ is the same, i.e.,
\begin{equation}
\Delta ^{2}\phi _{i}=\frac{1+e^{2r}}{4\left\vert z\right\vert ^{2}e^{2r}}.
\label{25}
\end{equation}%
It can be demonstrates that, in the case of the $\left\vert z\right\vert
^{2}=\sinh ^{2}r=\bar{n}_{\text{in}}/2$, $\Delta ^{2}\phi _{i}\,<1/\bar{n}_{%
\text{in}}$. It can surpass the SNL, but not reach the QCRB expressed by Eq.
(\ref{a0}). Particularly, in the case of the $r=0$, we can further obtain
the concise expressions of the CFIM for a coherent-light MZI with the $\phi
_{1}-\phi _{2}$ model
\begin{equation}
F^{c}|_{r=0}=2\left\vert z\right\vert ^{2}\left[
\begin{array}{cc}
\cos ^{2}\phi _{1} & 0 \\
0 & \cos ^{2}\phi _{2}%
\end{array}%
\right] .  \label{e3}
\end{equation}%
Comparing Eq. (\ref{e3}) with Eq. (\ref{9}), one can see that the CFIM
reduce to the QFIM when both $\phi _{1}$ and $\phi _{2}$ sit at the optimal
working points $\pm k\pi $. Therefore, for two-parameter estimation, the
optical phase sensitivity obtained by the double-port homodyne detection in
a coherent-light MZI can approach to the QCRB. Then, for the coherent-light
MZI with the $\phi _{1}-\phi _{2}$ model, the double homodyne detection is
the optimal measurement. However, for a squeezing-light MZI with the $\phi
_{1}-\phi _{2}$ model, although the SVS can further improve the phase
sensitivity, the double homodyne detection is not the optimal measurement
for such probe states as shown in the following.

Now we turn to investigate the performance of the double homodyne detection
explicitly and more exactly based on Eqs. (\ref{22}) by the numerical
analysis. In Fig. 3, we plot the phase sensitivity as a function of both
phase $\phi _{1}$ and $\phi _{2}$ for given values of parameters $z$ and $r$%
. It can be seen clearly from Fig. 3 that the value of one estimated phase
will affect the phase sensitivity of the other one. Figure 3 (a) indicates
that the phase sensitivity of the phase $\phi _{1}$ will always surpass the
SNL for its any values when the values of the $\phi _{2}$ around $\pm k\pi $
$(k=0,1,2\cdots )$. The converse is equally true as shown in Fig. 3 (b).
When the values of both $\phi _{1}$ and $\phi _{2}$ around zero, the total
phase sensitivity of both estimated phase will reach it optical case as
shown in Fig. 3 (c).

Particularly, in the case of $\left\vert z\right\vert ^{2}=\sinh ^{2}r=\bar{n%
}_{\text{in}}/2$, we can see from Fig. 4 that the phase sensitivity of the
phase $\phi _{i}$ not only can surpass the SNL, but also can surpass the HL
for a small range of phases. Different from the coherent-light MZI, we can
also see for the squeezing-light MZI with the double homodyne detection that
the optimal working points of the phases are not zero, but very close to
zero. Actually, by numerical analysis, it shows that the optimal working
points of the phases depend on these parameter of the input state. On the
other hand, comparing with Eq. (\ref{12}), we can demonstrate by numerical
analysis that the CRB of the double-port homodyne detection does not
saturate the QCRB. Therefore, for the squeezed-light MZI with the $\phi
_{1}-\phi _{2}$ model, the double homodyne detection is not the optimal
measurement. This may be because that the SVS as the probing state, the
measurement of $\left\langle \hat{X}\left( \theta \right) \right\rangle $
does not yield information about the estimated phase as shown in Eqs. (\ref%
{a1}) and (\ref{a2}).
\begin{figure}[tbp]
\centering\includegraphics[width=8cm]{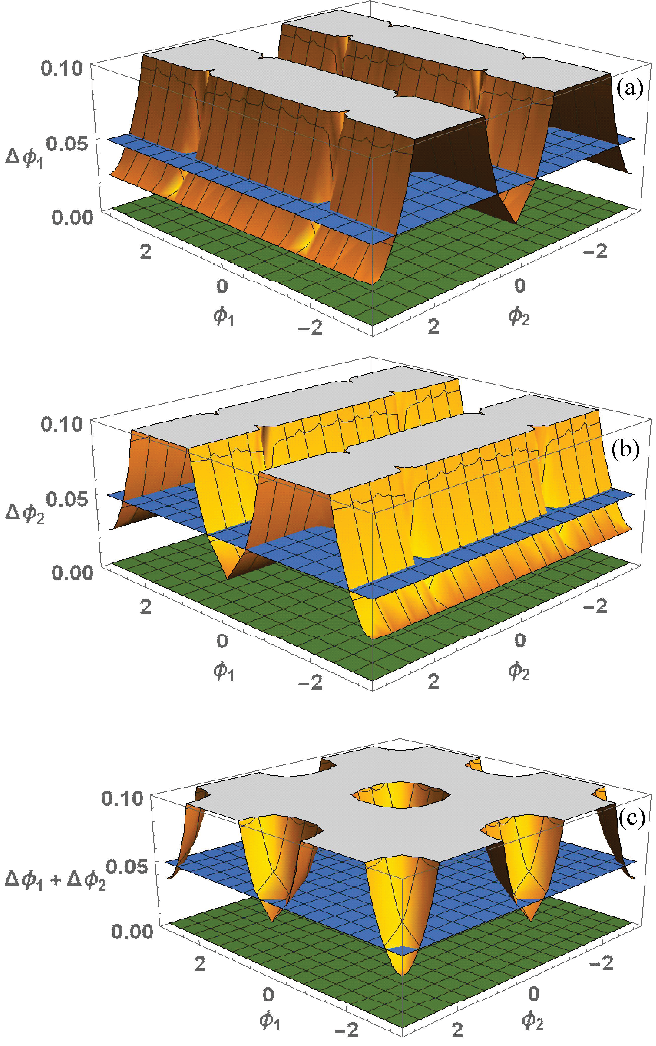}
\caption{(Color online) The phase sensitivity varies with both phases for
given $r=1.2$, and $\left\vert z\right\vert =20$. (a) The phase sensitivity
of the $\protect\phi _{1}$. (b) The phase sensitivity of the $\protect\phi %
_{2}$. (c) The total phase sensitivity of both $\protect\phi _{1}$ and $%
\protect\phi _{2}$. The blue plane represents the SNL, while the green plane
indicates the HL. }
\end{figure}
\begin{figure}[tbp]
\centering\includegraphics[width=8cm]{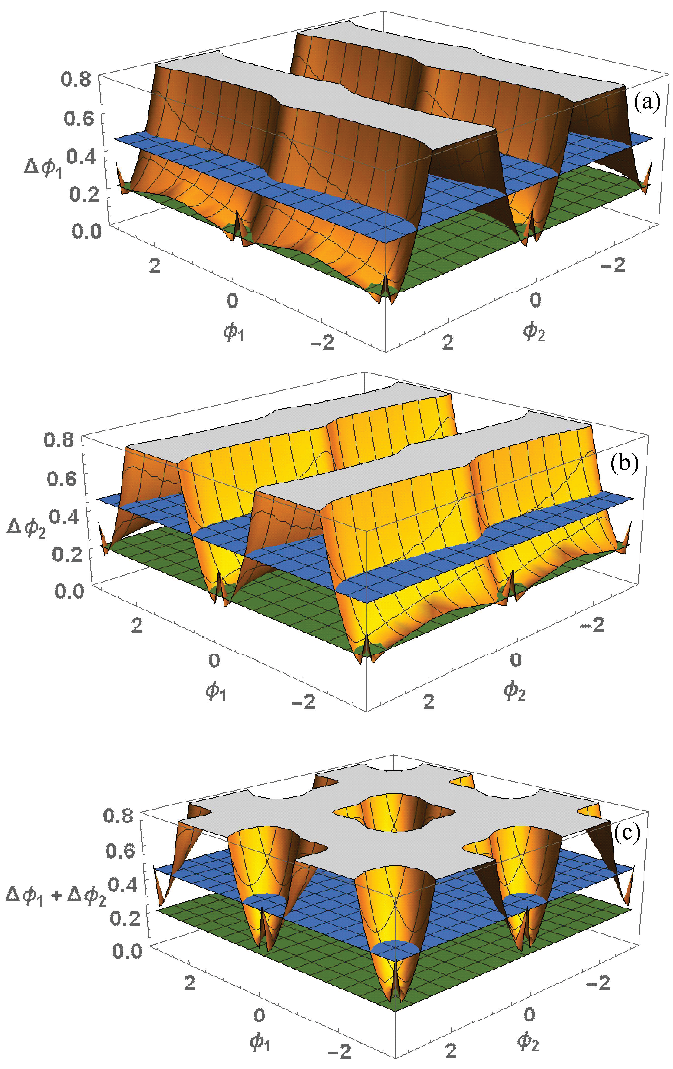}
\caption{(Color online) The phase sensitivity varies with both phases for
given $r=1.2$ and $\left\vert z\right\vert ^{2}=\sinh ^{2}r$. (a) The phase
sensitivity of the $\protect\phi _{1}$. (b) The phase sensitivity of the $%
\protect\phi _{2}$. (c) The total phase sensitivity of both estimated
phases. The blue plane represents the SNL, while the green plane indicates
the HL. }
\end{figure}

In order to see clearly the varies of the phase sensitivity with respected
to the phase, according to Fig. 3 (a) and Fig. 4 (a), we plot the phase
sensitivity as a function of one phase $\phi _{i}$ in Fig. 5 when the other
phase (for example, $\phi _{2}=0$) around zero. One see again that although
the optimal working points of the phases are not zero, they are very close
to zero. For a given squeezing parameter, the phase sensitivity of one phase
shift is basically unchanged throughout the phase space when the intensity
of the coherent state is large enough. In general, although the phase
sensitivity can not surpass the HL, it can still surpass the SNL as shown in
Fig. 5 (a). On the other hand, in the case of $\left\vert z\right\vert
^{2}=\sinh ^{2}r$, the phase sensitivity can surpass the HL for a small
range of the phase as shown in Fig. 5 (b).
\begin{figure}[tbp]
\centering\includegraphics[width=8cm]{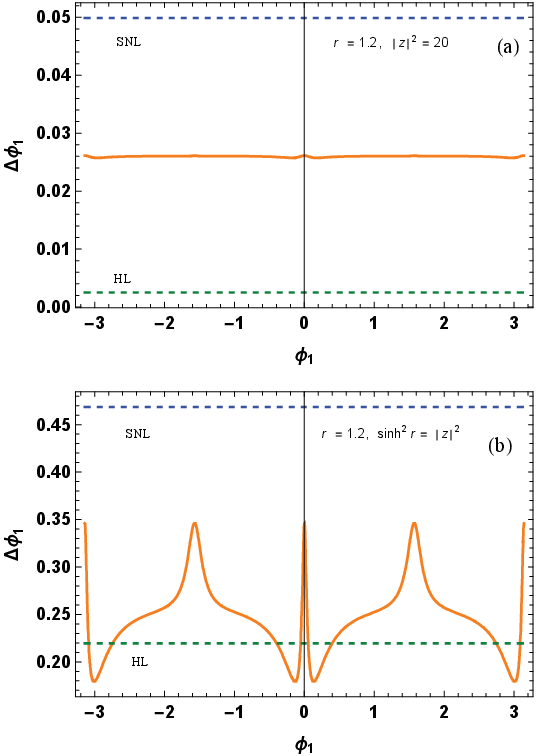}
\caption{(Color online) The phase sensitivity varies with the phase shift $%
\protect\phi _{1}$ for given $\protect\phi _{2}=0$, and $r=1.2$. (a) $%
\left\vert z\right\vert ^{2}=20$. (b) $\left\vert z\right\vert ^{2}=\sinh
^{2}r$. The blue dashed line represents the SNL, while the green dashed line
indicates the HL. }
\end{figure}

At the phase point $\phi _{1}=\phi _{2}=0$, we plot the phase sensitivity
versus the average total photon number of the input state in Fig. 6. We can
see that the SVS can further improve the phase sensitivity within a
constraint on the total photon number in the interferometer. In the case of $%
r=0$, the CRB saturates the QCRB as shown in Fig. 6 (a). For a given values
of the squeezing parameter $r$, when $\left\vert z\right\vert ^{2}\gg \sinh
^{2}r$, i.e., the intensity of the coherent state is large enough, the CRB
can asymptotically approach the QCRB too. Therefore, although the
double-port homodyne detection is not the optimal measurement, it is still a
quasi-optimal scheme.

\begin{figure}[tbp]
\centering\includegraphics[width=8cm]{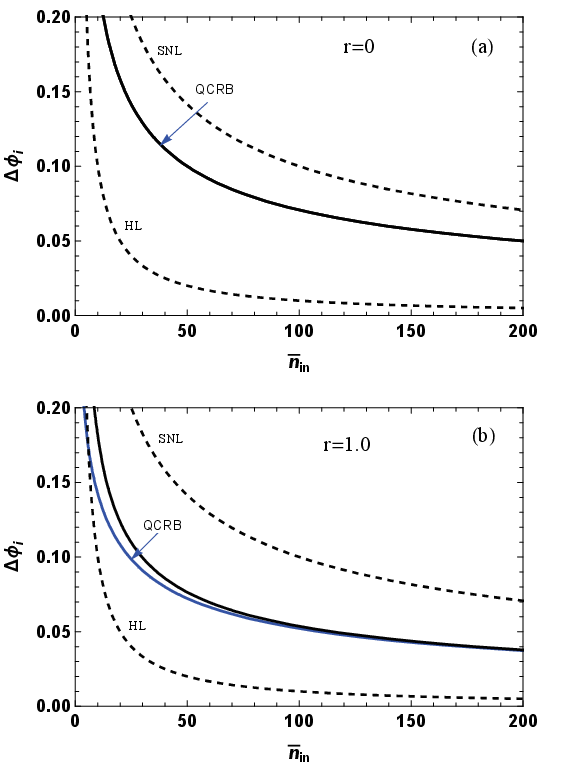}
\caption{(Color online) \ Phase sensitivity versus the average total photon
number of the input state for given values of the squeezing parameter $r$.
(a) $r=0\,$. (b) $r=1.0$. The upper black dashed line denotes the SNL, while
the below black dashed line represents the HL. line represents the HL.}
\end{figure}

On the other hand, for the phase shift generated only in one mode of the
MZI, the CRB defined by $\sqrt{1/F_{ii}^{c}}$ in the case of $\phi =0$ reads
\begin{equation}
\left( \Delta ^{2}\phi \right) _{\text{CRB}}=\frac{1}{\left\vert
z\right\vert ^{2}\left( 1+e^{2r}\right) },  \label{26}
\end{equation}%
which$\,$is consistent with that results in Refs. \cite{51,53}. Comparing
Eq. (\ref{26}) with Eq. (\ref{w1}), one can clearly see that the CRB of the
single-phase estimation also does not saturate the QCRB. It should be
pointed out that, for the phase shift generated only in one mode of the MZI,
the optimal working points of the phase is also not zero, but very close to
zero. Similarly, by numerical analysis, it shows that the optimal working
points of the phase also depends on these parameter of the input state.

\section*{Conclusions}

In summary, we have proposed a double-port homodyne detection to realized
two-parameter estimation in a squeezing-light MZI. According to the model
and basic theory of the general MZI with two-phase model, the measurement
signal of the homodyne detection is not only depended on the phase
difference, but also depended on the phase sum between the two arms of the
interferometer. Therefore, we can obtain the two phase shifts $\phi _{1}$
and $\phi _{2}$ simultaneously by the double-port homodyne detection. Via
the QFIM, our results show that the ultimate phase sensitivity, the QCRB, of
the estimated phase can be improved by the SVS, and can surpass the SNL,
even surpass the HL when half of the input intensity of the interferometer
is provided by the coherent state and half by the squeezed light. According
to the CFIM, the phase sensitivity of one estimated phase obtained by the
double-port homodyne detection can beat the SNL, and can be improved by the
SVS. Particularly, when the intensity of the coherent state is much larger
than that of the SVS, the phase sensitivity of the estimated phase will
asymptotically approach the QCRB. On the other hand, when half of the input
intensity of the interferometer is provided by the coherent state and half
by the squeezed light, the\ phase sensitivity of one estimated phase can
even surpass the HL for a small range of the phase. Although, for
two-parameter estimation in a squeezed-light MZI, the double-port homodyne
detection is not the optimal measurement, it is still a quasi-optimal
scheme. In addition, our method for two-parameter estimation with single MZI
may also be applicable to other probe states.

\section*{Declaration of Competing Interest}

The authors declare that they have no known competing financial interests or
personal relationships that could have appeared to influence the work
reported in this paper.

\section*{Acknowledgments}

This work was supported by the National Natural Science Foundation of China
(12104193).

\section*{Data availability.}

No data were generated or analyzed in the presented research.

\end{document}